\documentclass[prl,aps,twocolumn,showpacs]{revtex4}

\usepackage[dvips]{graphicx}
\usepackage{dcolumn}

\begin{document}



\title{Rate of photon production from hot hadronic matter}

\author{Kevin Lee Haglin}

\affiliation{Department of Physics, Astronomy \& Engineering
Science, St. Cloud State University, 720 Fourth Avenue South,
St. Cloud, MN 56301 \ USA}

\date{\today}

\begin{abstract}
Thermal photon emission rates from hot hadronic matter are studied 
to order $e^{2}g^{4}$, where $g$ indicates a strong-interaction
coupling constant.  Radiative decay of mesons, Compton and annihilation 
processes for hadrons, and bremsstrahlung reactions are all considered.
Compared to the standard rates from the literature, one finds two 
orders of magnitude increase for low photon energies
stemming mainly from bremsstrahlung and then a modest increase (factor
of 2) for intermediate and high energy photons owing to radiative
decays for a variety of mesons and from other reactions involving 
strangeness.  These results could have important consequences
for electromagnetic radiation studies at RHIC.
\end{abstract}


\maketitle

Photons carry undistorted detail on strong-interaction dynamics since 
their mean free paths in hot hadronic matter are much larger than
typical system sizes.  For this reason, photon measurements
in high energy nuclear collisions are deemed the best
thermometers of the highest energy densities reached 
there\cite{ef,es}.  While pions 
and protons scatter several times before leaving the collision zones, 
photons do not, and therefore
originate from the entire spacetime hadronic source, and in 
their asymptotic momentum configurations.
Quantitative studies of photon mean free paths in hadronic 
matter have indeed revealed $\lambda \simeq$
2$\times$10$^{3}$ fm for a temperature 200 
MeV\cite{kh_mfp}.  

Since production rates are increasing functions of temperature, the
inverse slope on a transverse momentum spectrum, for instance, 
is a good measure of the kinetic properties of the early 
(thermal) hadronic source.
And yet, photons are produced during the entire 
reaction.  Pre-equilibrium dynamics contribute, parton dynamics 
contribute, and then, hadron dynamics might be considered as a background
to the more sought-after quark gluon plasma 
contributions\cite{mtreview,cgkhreview}.  For
quantitative interpretation of photon measurements at the
Relativistic Heavy Ion Collider (RHIC), a complete determination
of all components, including the hadronic background, is 
clearly a prerequisite.

The early comparison of photon production from quark gluon plasma
versus hadronic matter by Kapusta {\it et al.\/}\cite{kls91}
revealed that the two phases have equal production rates at
fixed temperature.  Technically, the advancements
here were made possible through an application of hard-thermal-loop
resummation methods\cite{rp,bp} and effective hadronic field theory.
This physically appealing result
is however, somewhat discouraging in terms of allowing distinction
between the two phases using photons.  The equal-luminous property
of the two phases has survived even after next-to-leading-order (NLO) QCD
processes have been considered\cite{pa}.  NNLO parton processes have
even been analyzed\cite{pa00,pa00b,amy01}, and the new rate is 
still consistent with
the hadron rates.  The inevitable conclusion is that photon
production as a QGP diagnostic is not as definitive as was
originally hoped.   However, with the tremendous recent advancements
in the determination of QCD processes---to three-loop order, including
annihilation with scattering and also bremsstrahlung with
multiple scattering interferences\cite{lp,migdal}, it
behooves theory to put the hadronic formalism on a more equal
footing.  The purpose of this paper is to report on photon
production from hadronic matter computed at NNLO in
the strong-interaction coupling constant, namely, to
order $e^{2}g^{4}$.

The photon production rate to lowest order in the electromagnetic
coupling and to all orders in the strong coupling can be compactly
written as\cite{aw83,mt85,gk91}
\begin{eqnarray}
E_{\gamma}{d\/R\/\over\/d^{3}\/p_{\gamma}\/} & = & 
-{2\/g^{\mu\nu}\over\left(2\pi\right)^{3}}
\,{\rm\/Im}\Pi^{R}_{\mu\nu}(p_{\gamma})\,{1\over\/e^{\beta\/E_{\gamma}\/}-1},
\end{eqnarray}
where the retarded photon self-energy includes all possible 
hadron topologies---that is, to arbitrary loop order.  The imaginary 
part corresponds to 
emission and absorption, of which the former is relevant here.
Absorption was studied in Ref.~\cite{kh_mfp} and found to be negligible.   At 
the one-loop level, the imaginary
part corresponds to such lowest-order reactions as $\omega\to\pi\gamma$ 
and $\rho\to\pi\gamma$.  There is of course a long list of
other resonance decays to be considered, as discussed below.   Typically, 
in order to quantify production rates, a model for the strong interaction
(e.g. an effective Lagrangian) is
necessary.  However, the one-loop reactions have the
desirable feature that the rate can be computed
in a model independent way, requiring only the measured radiative
decay rates.  If such measurements are not available, then vector meson
dominance plus a chiral Lagrangian is an appropriate approach.

For the process $a\to\/1+\gamma$, where $a$ and 1 are mesons
with allowed quantum numbers and masses, the thermal rate can be
written as
\begin{eqnarray}
E_{\gamma}{d\/R\/\over\/d^{3}\/p_{\gamma}\/} & = & {{\cal\/N\/}m_{a}^{2}\,
\Gamma(a\to\/1\gamma)
\over 16\pi^{3}E_{\gamma}\/E_{0}}
\int\limits_{E_{\min}}^{\infty}\,dE_{a}\,f_{BE}(E_{a})
\nonumber\\
& \ & \times
\left[
1+f_{BE}(E_{a}-E_{\gamma})\right]\/,
\end{eqnarray}
where ${\cal\/N\/}$ is the spin and isospin degeneracy for
the parent, $E_{0}$ is the photon energy in the rest frame of the parent,
$E_{\min} = m_{a}(E_{\gamma}^{2}+E_{0}^{2})/2E_{\gamma}E_{0}$,
and $f_{BE}$ is the Bose-Einstein distribution to account
for quantum statistics and medium-enhancement for species 1.
If baryons were studied, e.g. $\Delta\to\/N\gamma$, then
Fermi-Dirac distributions would be appropriate and a minus
sign in the second factor of the integrand would suppress 
phase-space occupation for species 1.
The above integral is typically reported in the literature, and
then a numerical exercise ensues.  However, there is no need to leave
it here since the integral has a closed-form expression with the result
\cite{feynman_integral}
\begin{eqnarray}
E_{\gamma}{d\/R\/\over\/d^{3}\/p_{\gamma}} & = &
{{\cal\/N\/}m_{a}^{2}\,\Gamma(a\to\/1\gamma)
\over\/16\pi^{3}\,E_{\gamma}\,E_{0}}\,f_{BE}(E_{\gamma})
\nonumber\\
& \ & \times\/T\left[
\ln\left({f_{BE}(E_{\min}-E_{\gamma})\over\/f_{BE}(E_{\min})}\right)
-E_{\gamma}/T\right].
\label{my1loop}
\end{eqnarray}

\begin{figure}[t]
\begin{center}
\includegraphics[width=6.0cm,height=7.5cm]{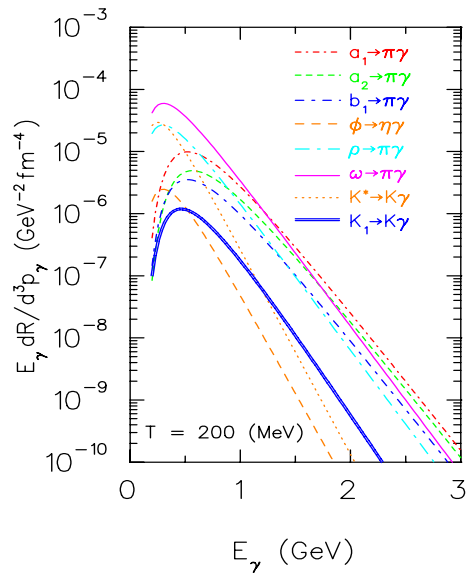}
\caption{Photon production rates from lowest-order reactions of
radiative meson decays at fixed temperature $T = 200$ MeV.}  
\label{rateI}
\end{center}
\end{figure}

Empirical radiative decay widths for several low-lying
resonances are available\cite{pdg} and
included in Table~\ref{gwidths}.  The resulting thermal decay rates
are displayed in Fig.~\ref{rateI}.  
\begin{table}
\caption{\label{gwidths}Empirical radiative decay widths
for low-lying resonances.} 
\begin{ruledtabular}
\begin{tabular}{lr}
Process & Width (keV)\\
\hline
$\rho(770)\to\pi\gamma$  & 117.9 \\
$\omega(783)\to\pi^{0}\gamma$  & 734.2 \\
$K^{*}(892)\to\/K\gamma$  & 83.5 \\
$\phi(1020)\to\eta\gamma$  & 55.3 \\
$b_{1}(1235)\to\pi\gamma$  & 227.2 \\
$a_{1}(1260)\to\pi\gamma$  & 640$\pm\/$240 \\
$a_{2}(1320)\to\pi\gamma$  & 286 \\
$K_{1}(1270)\to\/K\gamma$  & 70\footnote{Vector meson dominance was used to 
estimate this value.} \\
\end{tabular}
\end{ruledtabular}
\end{table}

Next-to-leading-order contributions enter at order $e^{2}g^{2}$
and come from two-loop self-energy
structures.   Coupling strength and phase space arguments
suggest that $\pi\rho\to\pi\gamma$, $\pi\pi\to\rho\gamma$ and 
$\rho\to\pi\pi\gamma$ will be important contributors.
The first channel, which is the Compton type process, will
be considered here while the second and third will be discussed
later in the higher orders, since they are essentially
the on-shell components of the full, higher order
contributions.  For the
Compton process, a particular interaction Lagrangian was assumed 
in Ref.~\cite{kls91} describing
$\pi$-$\rho$ dynamics, and then calibrated
to $\rho\to\pi\pi$ decay.  In terms of practical 
kinetic theory formalism, the rate for the general 
reaction $a+b\to\/1+\gamma$ is
\begin{eqnarray}
E_{\gamma}{d\/R\/\over\/d^{3}\/p_{\gamma}} & = &
{\cal\/N\/}\int
{d^{3}\/p_{a}\over(2\pi)^{3}\/2\/E_{a}}f_{a}
{d^{3}\/p_{b}\over(2\pi)^{3}\/2\/E_{b}}f_{b}
{d^{3}\/p_{1}\over(2\pi)^{3}\/2\/E_{1}}\tilde{f}_{1}
\nonumber\\
& \ & 
{1\over(2\pi)^{3}\/2\/}\,|\bar{\cal\/M\,}|^{2}
(2\pi)^{4}\/\delta^{4}\left(
p_{a}+p_{b}-p_{1}-p_{\gamma}
\right).\quad
\label{compton}
\end{eqnarray}

\begin{figure}[t]
\begin{center}
\includegraphics[width=6.0cm,height=7.5cm]{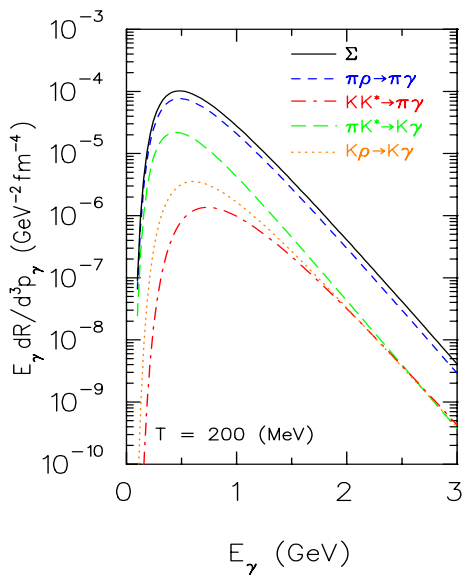}
\caption{Thermal photon production rate from important
channels at order $e^{2}g^{2}$.}
\label{rateII}
\end{center}
\end{figure}

A numerical task remains:  That is, to integrate Eq.~(\ref{compton}) over the
appropriate phase space.  The result for
$\pi\rho\to\pi\gamma$ is shown in Fig.~\ref{rateII}.

There are many other contributions at the same order with
roughly the same phase space properties.  
Certain strangeness reactions
turn out to be strong contributors, as can be seen from the
results plotted in Fig.~\ref{rateII}.  The coupling constants are
fitted to empirical hadronic decays where available, and estimated
using an SU\/(3) symmetry in a chiral Lagrangian  in the absence 
of measurements.
The strongest contributor involving strangeness seems to be
$\pi\/K^{*}\to\/K\gamma$, although there are certainly others of some
significance.  
Strangeness reactions of this type go beyond what has been done
before\cite{kh94}.
Also, these two-loop reactions and others are currently
being studied in a parallel investigation, broadened in scope to
explore the issue of form factors\cite{mcgillgroup}.

A remark is in order here regarding the role of the $a_{1}$ meson.
In the reaction $\pi\rho\to\pi\gamma$, the $a_{1}$ appears as 
an intermediate state, both in the $s$ channel and the $u$ channel.
Early estimates reported that the $s$ channel dominated all others,
even dominated $\pi$ exchange\cite{xsb}. 
Later studies using a hidden local symmetry in a
chiral Lagrangian were less clear regarding
the role of the $a_{1}$ since no unique rate could be established\cite{song}.
However, recent improvements
describing $a_{1}$ dynamics with particular concern for the
$D/S$ ratio in the $a_{1}\to\pi\rho$ scattering amplitude show
a diminished importance of the role of the $a_{1}$
for photon production\cite{simonandcharles}. 
In Fig.~\ref{rateI}, the $a_{1}$ radiative decay has been included while
in Fig.~\ref{rateII} only the $\pi$-exchange result for the Compton
process $\pi\rho\to\pi\gamma$ is shown.
This has the slight incomplete feature of ignoring the $u$ channel
$a_{1}$ exchange and
coherence effects, but at the same time has the advantage of removing any
possibility for double counting.

Motivated partly by the realization that the QCD processes
and the resulting photon rates have been computed with
bremsstrahlung included (and the hadron rates up to now have not),
one moves next to three-loop topologies.  Three-loop self-energy graphs
involve four on-shell hadron lines with a variety of
possible kinematics, plus the photon.  Some of these include 
processes of the type $a+b\to\/1+2+\gamma$:  hadronic elastic
scattering with bremsstrahlung.  Based on general principles alone,
these contributions are expected to be strong for low photon
energies and quickly become less important for increasing
photon energy.  Indeed, they
will even diverge as $E_{\gamma}\to\/0$.  When bremsstrahlung processes
are included, there is immediately a danger of double counting.
For example, $\pi\pi\to\rho\gamma$ and $\rho\to\pi\pi\gamma$ are
already part of the full $\pi\pi\to\pi\pi\gamma$ coherent calculation
with an intermediate $\rho$ meson.
Even though the two separate reactions were generated from
two-loop self-energy topologies (restricted, however, to on-shell
vector mesons), they must not be added on top of the bremsstrahlung
contributions.  That would be double counting.  The strategy taken here
is the following.  When two-loop
contributions are subsets of three-loop contributions, the separate two-loop
contributions will not be included.

A treatment of the bremsstrahlung mechanisms in terms
of complete phase space is a bit tedious, and so to first
assess their relative importance, a soft-photon approximation is
adopted here.  The rate is then\cite{hge93}
\begin{eqnarray}
E_{\gamma}{d\/R\/\over\/d^{3}\/p_{\gamma}} & = &
{\cal\/N\/}{T^{2}\over\/16\pi^{4}}\,\int\limits_{z_{\min}}^{\infty}\,
d\/z\/\,\lambda(z^{2}T^{2},m_{a}^{2},m_{b}^{2})\,K_{1}(z)
\nonumber\\
&\ & 
\times\left\lbrack
E_{\gamma}{d\sigma\/\over\/d^{3}\/p_{\gamma}}
 \right\rbrack
  {R_{2}(s_{2},m_{a}^{2},m_{b}^{2})\over
R_{2}(s,m_{a}^{2},m_{b}^{2})}\/,
\label{spa}
\end{eqnarray}
where $z_{\min}T = E_{\gamma} + \sqrt{M^{2}+E_{\gamma}^{2}}$, with
$M = \max(m_{a}+m_{b},m_{1}+m_{2})$, 
where the full squared amplitude is approximated as $|{\cal\/M\/}|^{2}$
$\approx$ $|{\cal\/M\/}_{0}|^{2}$($-J^{2}$), where
${\cal\/M\/}_{0}$ is the elastic scattering amplitude, $J^{\mu}$ is the
hadronic current including the fundamental charge $e$, 
$K_{1}$ is the modified Bessel function of order 1, and
finally, where the ratio of the two-body phase space for $s_{2} =
s - 2\sqrt{s}\/E_{\gamma}$ and $s$ respectively, is a
phase-space correction factor to minimize error at
finite $E_{\gamma}$.
\begin{figure}[t]
\begin{center}
\includegraphics[width=6.0cm,height=7.5cm]{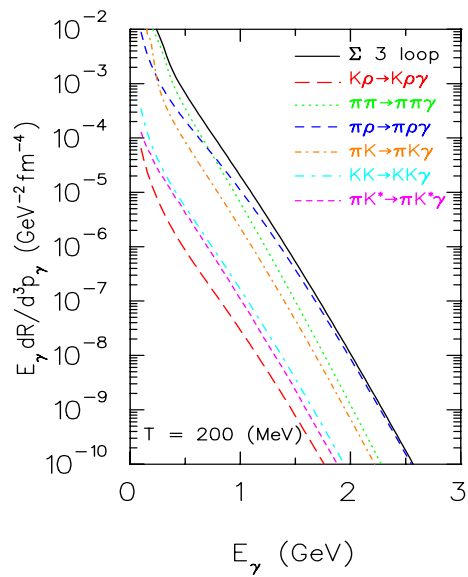}
\caption{Bremsstrahlung contributions (order $e^{2}g^{4}$) to thermal photon
production rate at $T$ = 200 MeV.}
\label{rateIII}
\end{center}
\end{figure}

Bremsstrahlung reactions are shown in Fig.~\ref{rateIII} for
pions, rho mesons, $K$ and $K^{*}$ mesons.  Interactions
are once again modeled with an SU(3) chiral Lagrangian with
coupling constants fitted by respecting hadronic phenomenology. 
As expected,
the partial rates are very strong at low photon
energy.  But the striking feature is the strength surviving
even for intermediate photon energies.

Two cautionary remarks must however be made.  First, Eq.~(\ref{spa})
does not include multiple scattering interference effects
of Landau-Pomeranchuk-Migdal, which will reduce the rates when
the photon energy is less than the inverse of the mean time
between strong interactions and second, the soft photon
approximation becomes less reliable with increasing photon
energy.  Still, one does not expect more than a factor of 2--4
suppression for $E_{\gamma}$ $\sim$ 0.5 GeV.

\begin{figure}[t]
\begin{center}
\includegraphics[width=6.0cm,height=7.5cm]{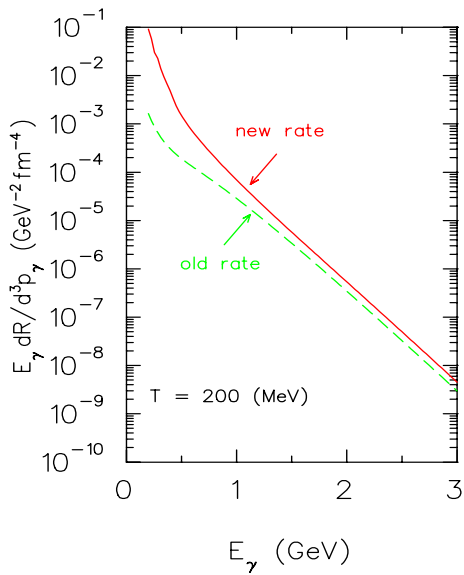}
\caption{Comparison of conventional hadronic rate to
order $e^{2}g^{2}$ (``old rate'' as parameterized by 
Nadeau {\it et al.\/}\cite{nad}) with 
an updated rate including additional radiative decays,
strangeness reactions, and bremsstrahlung (``new rate'') which goes to
order $e^{2}g^{4}$.  The old rate
includes $\pi\pi\to\rho\gamma$, $\rho\to\pi\pi\gamma$,
$\pi\rho\to\pi\gamma$ and $\omega\to\pi\gamma$.}
\label{ratecompare}
\end{center}
\end{figure}

Finally, the most useful comparison one can make regarding
the relative importance of the higher-order contributions is to add up 
all the partial rates from radiative decays (order $e^{2}$), 
Compton processes (order $e^{2}\/g^{2}$), and from bremsstrahlung 
(order $e^{2}g^{4}$)
and then compare to the conventional hadronic rate from the
literature (which includes $\omega\to\pi^{0}\gamma$,
$\pi\pi\to\rho\gamma$, $\rho\to\pi\pi\gamma$ and $\pi\rho\to\pi\gamma$).
This comparison is reported in Fig.~\ref{ratecompare}.
A factor of 2 increase is seen over most of the intermediate
and high energy region up to 3 GeV.  Below 1 GeV photon energy,
the increase is rather large, approaching two orders of magnitude
around 200 MeV photon energy.

One might summarize this study as follows.  Photon production rates
in hot hadronic matter have been investigated up to next-to-next-to-leading
order in the strong coupling.  This means contributions to order
$e^{2}g^{4}$ have been included, most notably, elastic
scattering of mesons with accompanying bremsstrahlung.  In addition,
strangeness reactions were included and found to be somewhat important.
Quantitatively, the new photon rate with strangeness and with
NNLO hadronic processes is two orders of magnitude greater
at low photon energies and a factor of approximately 2 greater at intermediate
and high energies as compared with the standard photon production
rate from the literature\cite{nad}.

Since multiple scattering interferences have not yet been included,
the precise increase at low photon energies reported here in the new
rate as compared with the old rate is not quantitatively
determined.
However, the general conclusions are expected to be robust.
It will therefore become extremely important
to assess such effects in future studies, as well, to incorporate
these new rates into dynamical or statistical models for
the purposes of comparison with experimental results at RHIC
and interpretation regarding QGP diagnostics.

{\bf Acknowledgments.}
I am grateful to Charles Gale and Simon Turbide for
useful discussions since they are carrying out a parallel
investigation, emphasizing form factor effects, strangeness 
contributions and axial vector mesons.

This work has been supported in part by the National
Science Foundation under grant number PHY-0098760.



\begin{thebibliography}{99}
\bibitem{feynman_integral}The Feynman parameterization $1\over{ab}$ = 
$\int\limits_{0}^{1}\,dx [ax+b(1-x)]^{-2}$ is useful for this task.
\end{thebibliography}
\end{document}